\documentclass[floatfix,aps,prb,eqsecnum,groupedaddress]{revtex4}
\usepackage[]{epsfig}

\newcommand{\be}{\begin{equation}}
\newcommand{\ee}{\end{equation}}
\newcommand{\bea}{\begin{eqnarray}}
\newcommand{\eea}{\end{eqnarray}}
\begin{document}
\title{Susceptibility 
  and Dzyaloshinskii-Moriya interaction in the Haldane gap compound NENP}
\author{Hai Huang}
\email{huanghai@physics.bu.edu}
\author{Ian Affleck}
\email{iaffleck@physics.ubc.ca}
\affiliation{Physics Department, Boston University, 590 Commonwealth Ave.,
Boston, MA02215 \\ 
Department of Physics and Astronomy, University of
British Columbia, Vancouver, BC, Canada, V6T 1Z1}
\date{\today}

\begin{abstract}
The Haldane gap material NENP exhibits anomalies in its Knight shift, 
far infrared absorption and field-dependent gaps, which have been 
explained using the staggered $g$-tensor that occurs due to the low 
crystal symmetry. We point out that the low-temperature susceptibility 
is also anomalous and that a consistent interpretation of all 
data may require consideration of the Dzyaloshinskii-Moriya interaction.
\end{abstract}
\maketitle
\section{Introduction}
Ni(C$_{2}$H$_{8}$N$_{2}$)$_{2}$NO$_{2}$ (ClO$_{4}$)(NENP),
is one of the best-studied quasi-one-dimensional antiferromagnets 
which exhibits a ``Haldane gap'' in its excitation spectrum since 
the atomic spins have S=1. The inter-chain coupling, $J'$, is estimated 
to be only $.0004J$, where $J\approx 48K$ is the intra-chain coupling and 
the disordered phase appears to persist down to zero temperature. 
Ignoring inter-chain couplings, the standard Hamiltonian for this 
system consists of Heisenberg exchange  plus crystal field terms:
\be H=\sum_j\{J\vec S_j\cdot \vec S_{j+1}+E^z(S^z_j)^2+E^x[(S^x_j)^2
-(S^y_j)^2]\}.
\label{Ham0}\ee
The crystal 
field interactions split the triplet magnon excitation into 3 separate modes 
at energies $13.6K$, $15.7K$ and $29K$.\cite{REGNAULT92} 

However, various anomalies appear in the finite field behavior of NENP. The 
low temperature susceptibility is much larger than expected from the 
measured gap anisotropy.\cite{AFFLECK90} 
The gap does not close at the Ising transition predicted to occur at 
a finite critical field.\cite{LU91} At low $T$ the Knight shift (local 
magnetic field at a nucleus) is much larger than expected.\cite{CHIBA91}
Production of a single magnon by far infrared absorption is observed even 
though this is expected to produce only zero wave-vector excitations 
and a single magnon has wave-vector near $\pi$.\cite{LU91} 

Chiba et al.\cite{CHIBA91} pointed out that the Knight shift
anomaly can 
be explained  by taking into account 
the staggered part of the gyromagnetic tensor. They observed that the 
local crystal structure near a magnetic Ni ion has principal 
axes which are rotated from the global crystal axes and 
that the local principal axes take two different orientations 
for even and odd sites along a chain. The $g$-tensor and also 
the crystal field Hamiltonian are expected to align with 
the local crystal symmetry. This implies that the $g$-tensor 
has a staggered component so that an applied uniform magnetic 
field leads to a small effective staggered field in addition 
to the uniform one. Because an antiferromagnet responds much 
more strongly, at low $T$, to a staggered field than to a 
uniform one, this leads to large effects at low $T$. By considering 
the direction of the staggered field, this 
theory is successful at explaining the various satellites of 
the proton Knight shift associated with the various inequivalent H-atoms 
in the unit cell.  

Mitra and Halperin\cite{MITRA94}observed that this staggered field also 
provides a natural explanation for the field-dependence of the gaps.
Since this staggered field is perpendicular to the uniform field 
it breaks the $Z_2$ symmetry that would otherwise be present 
and eliminates the finite field Ising transition. Using a 
mean field type approximation they attempted to fit  
the field-dependent gaps by the estimated staggered $g$-tensor.\cite{CHIBA91} 
Furthermore, 
because the field is staggered it halves the unit cell making 
wave-vectors $0$ and $\pi$ equivalent thus explaining 
the far infared adsorption anomaly. 

So far, no explanation has been offered, as far as we know, for
the anomalously large low $T$ susceptibility. Here we observe
that the staggered field also provides a natural explanation 
for this since the measured susceptibility then becomes 
a sum of uniform and staggered susceptibilities and the 
latter becomes quite large (but remains finite) at low $T$. However, 
we find that it is not possible to consistently fit the susceptibility data 
in terms of a staggered $g$-tensor alone. 

We also observe
that another important effect has been left out 
of previous explanations of these anomalies. This is the 
Dzyaloshinskii-Moriya (DM) antisymmetric exchange interaction
\cite{DZYALOSHINSKII58,MORIYA60},
\begin{equation}
H_{DM}=\sum_j\vec D_j\cdot (\vec S_j\times \vec S_{j+1}).
\end{equation}
The low crystal symmetry of NENP permits this interaction as well 
as the staggered $g$-tensor. A convenient way of treating 
a DM interaction is to remove it by a gauge transformation. It 
is possible to exactly eliminate it in favor of a small symmetric 
exchange interaction and a small perturbation to the 
crystal field Hamiltonian which just slightly change the 
magnon energies. However, the combination of a DM interaction 
and a magnetic field is less benign. The gauge transformation 
transforms the uniform field into a combination of uniform, slowly rotating
uniform and staggered effective fields.
Thus the effective staggered field 
has two sources which are potentially of the same order 
of magnitude.  

We calculate the susceptibility including the staggered $g$-tensor
and DM interaction using a Ginzburg-Landau (GL) mean field approach.
\cite{AFFLECK89}  We also do the calculation using 
a type of fermionic mean field theory.\cite{TSVELIK90}
Either approach allows quite good fitting of 
the susceptibility data at low $T$. A reliable determination 
of these parameters will probably await accurate numerical results 
on one-dimensional chains and more accurate experiments. 
It is possible that still other effects which we continue
to ignore such as the staggered crystal field interaction 
and inter-chain couplings are important. 
Nonetheless, we expect that our basic conclusion that 
the DM interaction and staggered $g$-tensor are of roughly
equal importance in explaining these anomalies will 
remain true. 

In the next section we review the crystal symmetry of NENP. Using 
this plus high-$T$ susceptibility measurements we estimate 
the uniform and staggered $g$-tensor. We also derive the most general form 
of the DM interaction allowed by symmetry. We then go on to 
discuss the low-$T$ susceptibilities of GL and fermion models in 
Sec. III.
In Sec. IV we comment on other types of experimental data 
and other theoretical approaches. 

\section{$g$-tensor and Dzyaloshinskii-Moriya interaction in NENP}
\begin{figure}[h]
\centerline{\epsfysize=0.30\textwidth{{\epsfbox{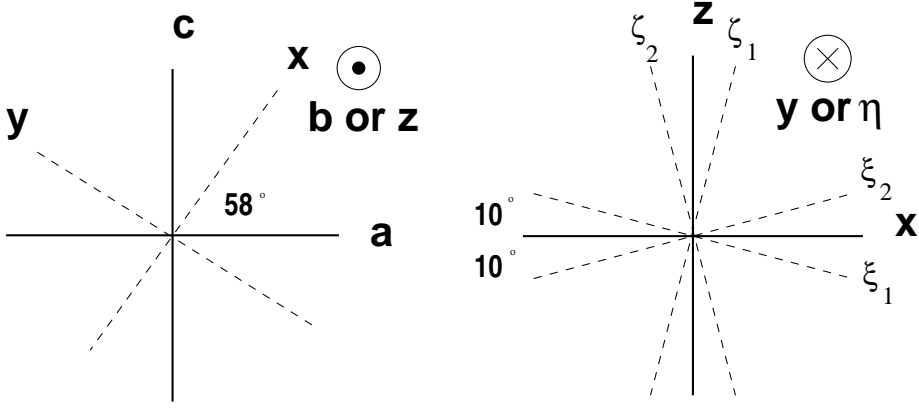}}}}
\caption{An illustration of the three co-ordinate systems. 
 1 and 2 refer to the  even and odd sites.}
\label{fig:axes}
\end{figure}

The ethylene-diamine molecule surrounding each magnetic Ni atom 
in NENP has an approximate orthorhombic symmetry with principal
axes  rotated relative to those defining the crystal space group. 
It is convenient to describe this rotation in two stages. Labeling 
the space group axes $(a,b,c)$ in the conventional way we first 
introduce a rotation matrix $\mathcal{R}_z$ which rotates 
by $58^{\circ}$ about the b-axis.  This defines a co-ordinate 
system which we label $(x,y,z)$.  (The chain axis, $b$ is 
identified with $z$.)  The components of the spin 
operators in this co-ordinate system ($S^a$) are related to those 
in the crystallographic system ($S^a{'}$) by:
\begin{equation}
\vec S=\mathcal{R}_z\vec S'.
\end{equation}
where $\mathcal{R}_{z}$ is a rotation about the z-axis by $(-\phi)$
($\phi \approx 58^{\circ}$):
\begin{equation} 
\mathcal{R}_{z}=
\left( \begin{array}{ccc}
\cos{\phi} & \sin{\phi} & 0 \\
-\sin{\phi} & \cos{\phi} & 0 \\
0 & 0 & 1
\end{array} \right ),
\end{equation}
A further rotation by $\pm \theta$ ($\theta \approx 10^{\circ}$)
about the y-axis, $\mathcal{R}_y^{\pm}$, depending on sites:
\begin{equation} 
\mathcal{R}_{y}^{\pm}=
\left( \begin{array}{ccc}
\cos{\theta} & 0 & \pm \sin{\theta} \\
0 & 1 & 0 \\
\mp \sin{\theta} & 0 &  \cos{\theta}  
\end{array} \right ).
\end{equation}
 defines the 
local symmetry axes around a Ni site, $(\xi,\zeta ,\eta )$.
  The $+$ or $-$ sign occurs 
for even or odd sites along a Ni chain.  (See Fig. \ref{fig:axes}.) 
We label the corresponding spin components $\vec S''$,
\begin{equation}
\vec S''=\mathcal{R}_y^{\pm}\vec S.
\end{equation}
The ordinary exchange interaction in NENP is generally 
assumed to be of the Heisenberg form
\begin{equation}
H_{ex}=J\sum_j\vec S_j\cdot \vec S_{j+1}.\end{equation}
However, there 
is an important symmetry breaking in the crystal field 
Hamiltonian. This is expected to be diagonal in 
the $\vec S''$ co-ordinate system, as in Eq. (\ref{Ham0}),
\begin{equation}
H_{CF}=\sum_{j}\{E^{z}(S^z_j{''})^2+E^{x}[(S^x_j{''})^2-(S^y_j{''})^2]\}.
\end{equation}
Transforming to the $\vec S$ coordinate system, the 
crystal field Hamiltonian has a diagonal uniform part 
and a small off-diagonal staggered part. We will assume
that the staggered part can be ignored in what follows.
The uniform diagonal part could then be fit 
to the observed magnon gaps.  This 
implies that $E^z>>E^x>0$.

Another important source of anisotropy, when a magnetic 
field is applied, is the Land\'e $g$-tensor, $\mathbf{g}$.  
The Zeeman term in the Hamiltonian is written:
\begin{equation}
H_Z=\mu_B\sum_j \vec h \cdot \mathbf{g}_j \vec S_j.
\end{equation}
The $g$-tensor is assumed to be diagonal in the $(\xi ,\zeta ,\eta )$ basis:
\begin{equation} 
\mathbf{g}_{(\xi, \eta, \zeta)}=
\left( \begin{array}{ccc}
g_{\xi} & 0 & 0 \\
0 & g_{\eta} & 0 \\
0 & 0 & g_{\zeta}
\end{array} \right ).
\end{equation}
The uniform and staggered $g$-tensors in the 
$(x,y,z)$ co-ordinate system  
are given by:
\begin{eqnarray}
\mathbf{g}_{(x, y, z)}&=&\mathbf{g}^{u}+\mathbf{g}^{s} \nonumber \\
                      &=&(\mathcal{R}_{y}^\pm )
                         \mathbf{g}_{(\xi, \eta, \zeta)}
                         (\mathcal{R}_{y}^\pm )^{-1},
\label{g_xyz}
\end{eqnarray}
where
\begin{equation}
\mathbf{g}^{u}=\left( \begin{array}{ccc}
g_{x} & 0 & 0 \\
0 & g_{y} & 0 \\
0 & 0 & g_{z}
\end{array} \right ) 
=\left( \begin{array}{ccc}
g_{\xi}\cos{\theta}^{2}+g_{\zeta}\sin{\theta}^{2} & 0 & 0 \\
0 & g_{\eta} & 0 \\
0 & 0 & g_{\xi}\sin{\theta}^{2}+g_{\zeta}\cos{\theta}^{2}
\end{array} \right )
\label{gu}
\end{equation}
and
\begin{equation} 
\mathbf{g}^{s}=
\left( \begin{array}{ccc}
0 & 0 & (g_{\xi}-g_{\zeta})\sin{\theta}\cos{\theta}  \\
0 & 0 & 0 \\
(g_{\xi}-g_{\zeta})\sin{\theta}\cos{\theta} & 0 & 0
\end{array} \right ).
\label{gs}
\end{equation}
The gyromagnetic tensor in the crystallographic coordinate system 
(a, b, c) can be written as
\begin{equation}
\mathbf{g}_{(a, c, b)}=(\mathcal{R}_{z}\mathcal{R}_{y}^\pm)
                       \mathbf{g}_{(\xi, \eta, \zeta)}
                       (\mathcal{R}_{z}\mathcal{R}_{y}^\pm )^{-1}.
\end{equation}

\subsection{Dzyaloshinskii-Moriya interaction  }
As discussed by Dzyaloshinskii\cite{DZYALOSHINSKII58} and 
Moriya\cite{MORIYA60}, 
an additional  exchange interaction term can appear in  
the Hamiltonian which is anti-symmetric under interchanging 
the two sites,  the DM interaction
\begin{equation}
H_{DM}=\sum_{j}\vec{D}_{j}\cdot(\vec{S}_{j}\times\vec{S}_{j+1}).
\end{equation}
The possible values of the DM vectors $\vec{D}_{j}$ can be limited by
considering crystal symmetries of NENP, which at low temperatures are 
given by the space group Pn$2_{1}$a\cite{MEYER82,MITRA94}. First, the compound
is invariant under a translation along the $\hat{b}$ (or $\hat{z}$) by 
two sites. This means the DM vectors are the same among the even (or odd) 
links. Second, the crystal structure is invariant under the combined 
operation of one site translation along the chain ($\hat{b}$) direction and a 
$180^{\circ}$ rotation around $\hat{b}$. The operation acts as 
$S_{j}^{a,c} \to -S_{j+1}^{a,c}$, $S_{j}^{b} \to S_{j+1}^{b}$.
This implies that $D_{a,j}$ and $D_{c,j}$ are staggered, 
$\propto (-1)^j$ while $D_{b,j}$ is uniform. 
The other symmetry operations  relate sites in one chain to sites in
the others, so there are no further restrictions on the intra-chain 
DM vectors. 

A nearest neighbor DM interaction in one dimension 
can always be eliminated by a redefinition of 
the spin operators which varies from site to site (i.e. a 
gauge transformation). Let us suppose that the symmetric exchange
interaction is $SO(3)$ invariant.  Then, choosing co-ordinates 
so that $\vec D\propto \hat z$,  we may write the combined 
symmetric and anti-symmetric exchange interactions as:
\begin{equation}
H_{ex}=\sum_j\{ [(J+iD_j)S^+_jS^-_{j+1}+h.c.]+JS^z_jS^z_{j+1}\}
\end{equation}
We may always transform this into a purely parity-symmetric exchange 
interaction:
\begin{equation}
H_{ex}=\sum_j[\sqrt{J^2+D^2}(S^x_jS^x_{j+1}+S^y_jS^y_{j+1})
+JS^z_jS^z_{j+1}],\end{equation}
by a gauge transformation:
\begin{equation}
S^-_j\to S^-_je^{i\alpha_j}.\label{gt}\end{equation}
 When 
$D_j=(-1)^jD$, the required gauge transformation simply 
alternates from site to site:
\begin{equation}
\alpha_j=(-1)^j (1/2) \tan^{-1}(D/J).\end{equation}
On the other hand, for a uniform $D_j=D$,
\begin{equation}
\alpha_j=\alpha  \cdot j,
\label{alpha.j}
\end{equation}
where
\begin{equation}
\alpha=\tan^{-1}(D/J).\end{equation}
This gauge transformation introduces a small xxz anisotropy 
into the symmetric exchange interaction.  Its effects 
on the crystal field Hamiltonian must also be considered. 
If we write this, in general, as:
\begin{equation}
H_{CF}=\sum_{j,a,b}S^a_jE_j^{ab}S^b_j,\end{equation}
then the effect of the gauge transformation is:
\begin{equation}
E_j\to {\cal R}(\alpha_j)E_j{\cal R}^{-1}(\alpha_j),\end{equation}
where ${\cal R}(\alpha_j)$ is the rotation matrix which effects 
the gauge transformation of Eq. (\ref{gt}):
\begin{equation}
{\cal R}(\alpha_j)=\left( \begin{array}{ccc}
\cos \alpha_j & \sin\alpha_j& 0\\
-\sin \alpha_j & \cos \alpha_j &0\\
0&0&1\end{array}\right).
\label{R(alpha_j)}
\end{equation}
Thus the principal axes of the crystal field Hamiltonian 
are rotated from site to site while the eigenvalues 
remain the same. 
For an alternating DM interaction, this is an 
alternating rotation which would introduce an 
alternating term in the crystal field Hamiltonian. 
As discussed above, such a term is expected to already 
be present, before the gauge transformation. 
For a uniform DM interaction, the transformed 
$E$-tensor in the crystal field Hamiltonian rotates 
steadily along the chain. We will assume these small
effects can be ignored.

The combination of a DM interaction and an applied 
field leads to more important effects. Upon performing
the gauge transformation, the $g$-tensor at site $j$ is 
transformed as:
\begin{equation}
g_j\to g_j{\cal R}^{-1}(\alpha_j).\end{equation}
In the case of a staggered DM interaction, this leads to an
alternating term in the $g$-tensor even if it  was not present before, 
thus adding to the effective staggered field.  A uniform DM 
interaction leads to a rotating effective magnetic field.  
Both staggered and uniform DM interactions can be readily treated 
using field theory methods. They appear to be
approximately as important as the staggered field in explaining
the various anomalies mentioned in Sec. I. We show that it is possible
to fit the susceptibility data quite well by taking into account the 
staggered and uniform DM interactions.

Since the DM interaction contributes  the 
same order of magnitude 
to the effective staggered field as the staggered gyromagnetic 
tensor\cite{OSHIKAWA97,AFFLECK99}, we have to combine them together. For small 
$\mathbf{g}^{s}$ and ($\frac{D}{J}$), and an arbitrary 
direction for the staggered DM vector $\vec{D}^{s}=(D_{x}, D_{y}, 0)$, 
the staggered field can be approximated as
\begin{equation}
\vec{h}^{s}\approx \mathbf{g}^{s}\vec{h}+(\frac{1}{2J})\vec{D}^{s} \times 
                   \mathbf{g^{u}}\vec{h}\equiv \mathbf{A}\vec h,
\label{hstag}\end{equation}
which is just the sum of two contributions \cite{OSHIKAWA97,AFFLECK99}. 
Here we have introduced another matrix $\mathbf{A}$ 
relating the total effective staggered field, $\vec h^s$
to the original laboratory field $\vec h$.  Note 
that $|\mathbf{A}|<<1$.

Thus after making the gauge transformation and discarding 
terms which we expect to be unimportant, the Hamiltonian 
can be written, in the $(x,y,z)$ ($\vec S$) co-ordinate system:
\begin{equation}
H=\sum_{j}\{J\vec{S}_{j}\cdot\vec{S}_{j+1}+E^z(S_{j}^{z})^{2}
+E^x[(S_{j}^{x})^{2}-(S_{j}^{y})^{2}]
-\mu_B\vec h \cdot [\mathbf{g^u} {\cal R}(\alpha\cdot j)] \vec S_j
-(-1)^j\mu_B\vec h^s \cdot {\cal R}(\alpha\cdot j) \vec S_j \},
\label{Ham}\end{equation}
where ${\cal R}(\alpha\cdot j)$ is defined in Eqs. (\ref{alpha.j}) 
and (\ref{R(alpha_j)}).  
The values of $J\approx 44K$ and $E^z\approx 8K$ 
have been determined from fitting the magnon gaps to numerical 
simulations.\cite{SORENSEN94, GOLINELLI92} 
$E^x\approx 0.4K$ is extracted by the best fit of experimental data 
to the six-spin-ring model calculation.\cite{LU91} 

\section{Susceptibility}
\subsection{Mean Field Results}
In the large-s approximation, the Heisenberg spin chain is equivalent to 
a field theory, the O(3) non-linear $\sigma$-model 
(see, e.g., Ref. \onlinecite{AFFLECK98} and Ref. \onlinecite{FRADKIN91}). 
The Hamiltonian of this model is given by 
\begin{equation}
H=(\frac{v}{2}) \int dz\left[g
\left.\vec{l}\,\right.^2
+\frac{1}{g}\left(\frac{\partial 
\vec{\phi}}{\partial z}\right)^{2}\right] 
\ \ \ (\left.{\vec{\phi}}\,\right.^{2}=1) \ \ ,
\end{equation}
where 
\begin{equation}
\vec{l}=\frac{1}{vg}\vec{\phi}\times\frac{\partial\vec{\phi}}{\partial t}. 
\end{equation}
The coupling constant $g$ and the magnon velocity take the values,
at $s \to \infty$,
\begin{equation}
g=\frac{2}{s}, \ \ v=2Js.
\end{equation}
The original spin operators are expressed in terms of the field $\vec{\phi}$
and the spin density, $\vec{l}$ as
\begin{equation}
\vec{S_{j}}\approx (-1)^{j}s\vec{\phi_{j}}+\vec{l_{j}}.
\end{equation}
If we relax the constraint of O(3) non-linear $\sigma$-model
and add a repulsive $\phi^{4}$ interaction, which is treated perturbatively, 
a much simpler theory can be 
obtained.
Including anisotropic terms, we then phenomenologically model the low-lying
excitations via the following bosonic quantum field theory
\cite{AFFLECK90,AFFLECK91}, which we refer to as the Ginzburg-Landau
model (GL model):
\begin{equation}
H=\int dz\left\{\sum_{i}\left[\frac{v}{2}\Pi_i^{2}
+\frac{v}{2}(\frac{\partial {\phi_i}}{\partial z})^{2}
+\frac{\Delta_{i}^{2}}{2v}\phi_{i}^{2}\right]
-\mu_{B}\sum_{iklmn} h_{i} {\mathbf g}^{u}_{ik} {\cal R}_{kl}(\alpha \cdot z)
\epsilon_{lmn}\phi_{m} \Pi_{n}
-\mu_{B}\sum_{ik}h^{s}_{i} {\cal R}_{ik}(\alpha \cdot z) \rho_{k}\phi_{k}
+\lambda \left.\vec{\phi}\,\right.^4
\right\}.
\label{GLM}
\end{equation}
Here $\epsilon_{ijk}$ is the anti-symmetric tensor with $\epsilon_{123}=1$.
We assume the gaps, normalization factors and velocity are:
\begin{eqnarray}
\Delta_{x}=15.7 K, \ \ \Delta_{y}&=&13.6 K, \ \ \Delta_{z}=29 K; \nonumber \\
\rho_{x}=\rho_{y}=1.08, \ \ \rho_{z}&=&1.2; \ \ v=120K; \ \ \lambda=3.7K.
\label{par} 
\end{eqnarray}
(We use units where the spacing between neighboring Ni ions 
along the chains is 1.)  The gaps are from neutron scattering 
experiments\cite{RENARD87} and 
normalization factors and velocity are from numerical 
simulations.\cite{SORENSEN94} 

Tsvelik proposed a fermionic field theory model\cite{TSVELIK90}
for NENP. We can easily include uniform DM interaction into this model. 
But there is some problem for staggered effective field (including staggered
$g$-tensor and staggered DM interaction). The staggered components of the
spin operators have a very complicated representation as the product of three
Ising order (and disorder) parameter fields. Consequently, there appears to
be no simple method for treating a staggered field in this model. If only
uniform DM interaction is taken into account, the Hamiltonian can be modified
as  
\begin{equation}
H=\int dz \left[\sum_k(i\bar{\chi}_{k}\gamma_{1}\partial_{z}\chi_{k}
            +\Delta_{k}\bar{\chi}_{k}\chi_{k})
            -\mu_{B}\sum_{klmnp}ih_{k}{\mathbf g}^{u}_{kl}{\cal R}_{lm}
            (\alpha\cdot z)\epsilon_{mnp}\bar{\chi}_{n}
            \gamma_{0}\chi_{p}\right],
\label{Tsvelik}
\end{equation}
where $\chi_{k}$ is two-component Majorana fermion field
\begin{equation} 
\chi_{k}=\left( \begin{array}{ccc}
\chi_{+,k}\\
\chi_{-,k}
\end{array} \right ) \ \ \ \ (k=1, 2, 3) \ \ ,
\end{equation}
the sign + (-) corresponds to the right (left) movers and
$\bar{\chi}=\chi^{T}\gamma_{0}$. $\gamma_{\mu}$ ($\mu=0, 1$)
are chosen as $\gamma_{0}=\sigma_{x}$, $\gamma_{1}=i\sigma_{y}$.
The advantage of this model is that the field-shifted gaps agree better 
with neutron scattering experiments than those of the bosonic model of 
Eq. (\ref{GLM}). 

\subsection{Isotropic Susceptibility: uniform and staggered}
In this subsection, we will forget about the DM interaction and crystal 
field terms, and discuss the uniform and staggered susceptibilities of 
isotropic Heisenberg spin-1 chain. In the isotropic case, when an external 
uniform magnetic field is applied to the system, we assume that 
the uniform $g$-tensor is also isotropic. We set $g \mu_{B}=1$ in this
subsection only.  

The Hamiltonian for this case is given by
\begin{equation}
H=\sum_{j}[J\vec{S}_{j}\cdot\vec{S}_{j+1}-\vec h \cdot \vec S_j].
\label{Ham-U}
\end{equation}

Assuming the excitations are non-interacting quasi-particles
and no gap anisotropy, the uniform zero-field susceptibility per spin 
is 
then 
\begin{equation}
\chi_{u}=\frac{1}{T}<(S_{z})^{2}>
        =\frac{1}{T}<(N_{+}-N_{-})^{2}>,
\end{equation}
where $N_{\pm}$ are total numbers of quasi-particles with $S_{z}=\pm1$.\\ 
Since $N_{+}$ and $N_{-}$ are equal in the ground state,
\begin{equation}
\chi_{u}=\frac{2}{T}(<N_{+}^{2}>-<N_{+}>^{2}).
\end{equation}
Write
\begin{equation} 
N_{+}=\sum_{k}N_{+,k},
\end{equation}
we get
\begin{equation}
\chi_{u}=\frac{2}{T}\sum_{k}(<N_{+,k}^{2}>-<N_{+,k}>^{2}).
\end{equation}
So for boson and fermion distributions, we have different 
susceptibilities per unit length as follows:
\begin{equation}
\chi_{u, B}=\frac{2}{T} \int \frac{dk}{2\pi} \frac{e^{\sqrt
{\Delta^2+v^2k^2}/T}}{(e^{\sqrt{\Delta^2+v^2k^2}/T}-1)^2}, 
\end{equation}
\begin{equation}
\chi_{u, F}=\frac{2}{T} \int \frac{dk}{2\pi} \frac{e^{\sqrt
{\Delta^2+v^2k^2}/T}}{(e^{\sqrt{\Delta^2+v^2k^2}/T}+1)^2},
\end{equation}
where the gap is 0.4107J\cite{WHITE92, WHITE93, SORENSEN93}
and the velocity is 2.5J.\cite{SORENSEN93}

\begin{figure}[h]
\centerline{
\epsfysize=0.50\textwidth{{\epsfbox{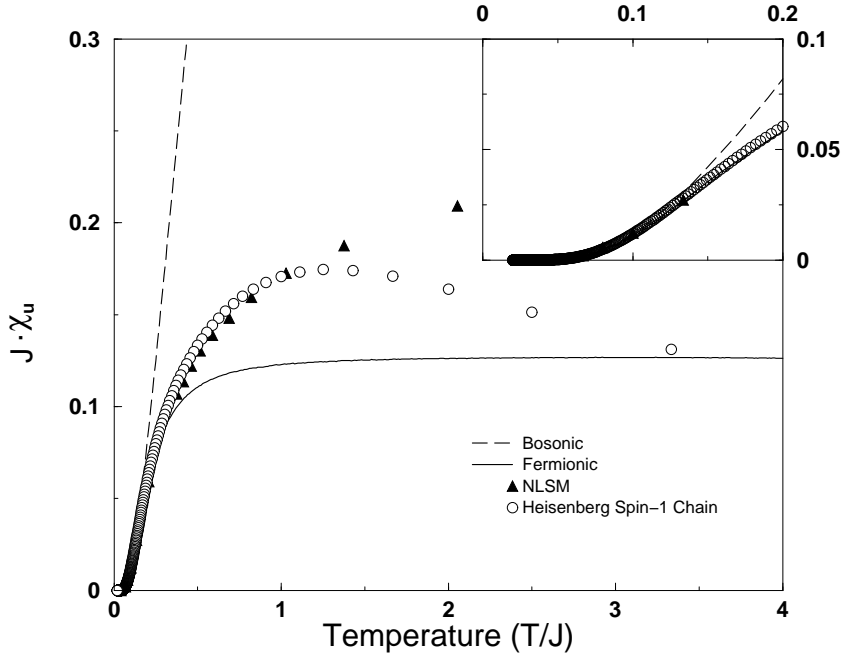}}}
}
\caption{Comparison of isotropic uniform susceptibilities of different models.}
\label{compare4}
\end{figure}

We plot isotropic boson, fermion, Heisenberg spin-1 chain (transfer-matrix 
renormalization-group method)\cite{XIANG98}, and non-linear 
$\sigma$-model results\cite{KONIK} in Fig. \ref{compare4}.
We see boson model result is consistent with Heisenberg spin-1 chain result
below around 0.2J, fermion model below roughly 0.5J, non-linear 
sigma model result is the best, below roughly 1.5J.

\begin{figure}[h]
\centerline{
\epsfysize=0.50\textwidth{{\epsfbox{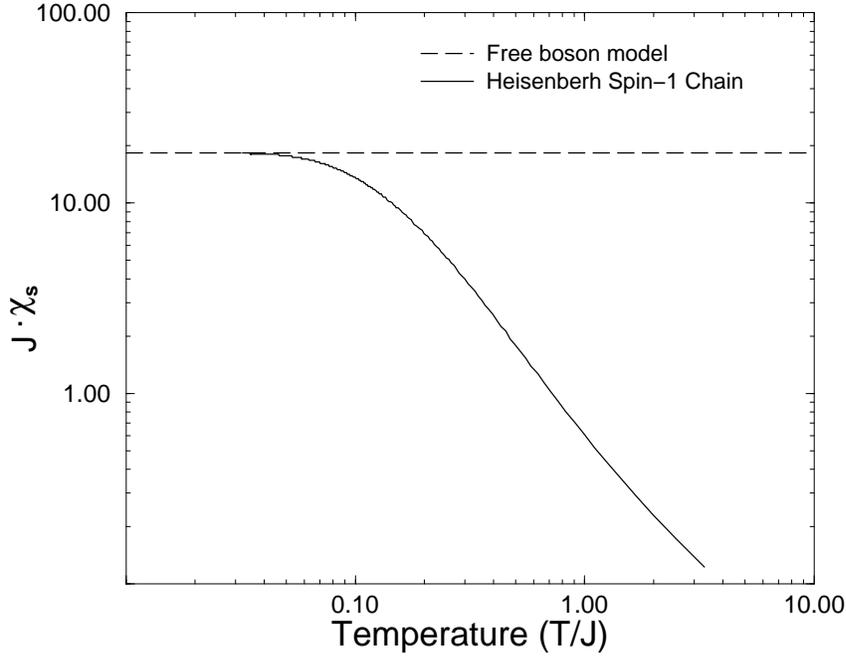}}}
}
\caption{Comparison of isotropic staggered susceptibilities of 
different models.}
\label{IsoStagg}
\end{figure}

When a staggered magnetic field is applied to the system, the Hamiltonian 
is
\begin{equation}
H=\sum_{j}[J\vec{S}_{j}\cdot\vec{S}_{j+1}-(-1)^j \vec h \cdot \vec S_j].
\label{Ham-S}
\end{equation}
For the free boson case, the staggered susceptibility is
\begin{equation}
J \cdot \chi_{s, B}=\frac{\rho^{2}(v/J)}{(\Delta/J)^2}=14.8\rho^{2},
\end{equation}
where $\rho$ is the wave function renormalization of bosonic field.
We choose $\rho=1.11$ to fit the low temperature results of Heisenberg
spin-1 chain (TMRG).\cite{XIANG98} They are plotted in Fig. \ref{IsoStagg}.
We see they are consistent up to around 0.1J.
We also notice this $\rho$ is consistent with the average value got from 
the numerical simulations of equal-time correlation function
\cite{SORENSEN94} (in Ref. \onlinecite{SORENSEN94}, $g_{i}=\rho_{i}^2$):
\begin{equation}
\bar{\rho}=\frac{1}{3}(\rho_{x}+\rho_{y}+\rho_{z})=1.12.
\end{equation}

\subsection{Susceptibility of GL Model: uniform and staggered}
Susceptibility data has been published for an applied field along 
the crystallographic $a$, $b$ or $c$ axis, which is anisotropic.
We assume that what is measured is 
$\partial^2F/\partial h_a^2|_0$ etc. where $F$ is the free energy 
and $h_a$ the field component in the $a$-direction.  
From Eqs. (\ref{Ham}) and (\ref{hstag}) we get in the (x, y, z)
co-ordinate system:
\begin{eqnarray}
\chi^{ik}
=\frac{1}{L}{\partial^2 F\over \partial h_i\partial h_k}
&=&\mu_B^2\sum_{lm}
[\mathbf{g}^u_{il}\mathbf{g}^u_{km}
(\frac{1}{L}\sum_{j_1 j_2,np}\int_0^\beta d\tau {\cal R}_{ln}
(\alpha \cdot j_{1}){\cal R}_{mp}(\alpha \cdot j_{2})
<S^n_{j_1}(\tau )S^p_{j_2}(0)>) \nonumber \\
&&+\mathbf{A}_{li}\mathbf{A}_{mk}
(\frac{1}{L}\sum_{j_1 j_2,np}(-1)^{j_1-j_2}
\int_0^\beta d\tau {\cal R}_{ln}(\alpha \cdot j_{1})
{\cal R}_{mp}(\alpha \cdot j_{2})<S^n_{j_1}(\tau )S^p_{j_2}(0)>)].
\label{sus}
\end{eqnarray}
Defining the reduced susceptibility $\bar{\chi}^{ik}$(q) 
at arbitrary momentum q:
\begin{equation}
\bar{\chi}^{ik}(q)=\frac{1}{L}\sum_{j_1 j_2}\int_0^\beta d\tau 
e^{q(j_{1}-j_{2})}<S^i_{j_1}(\tau )S^k_{j_2}(0)>,
\label{sus_bar}
\end{equation}
then plugging the explicit form of $\cal R$ 
matrix into Eq. (\ref{sus})
and use the translation invariance of spin correlation function,
$\chi^{ik}$ can be written as
\begin{equation}
\chi^{ik}=\mu_B^2\sum_{l}\left\{\mathbf{g}^u_{il}\mathbf{g}^u_{kl}
\left[\frac{1-\delta_{l,z}}{2}\bar{\chi}^{ll}(\alpha)+
\delta_{l,z}\bar{\chi}^{ll}(0)\right]
+\mathbf{A}_{li}\mathbf{A}_{lk}\left[\frac{1-\delta_{l,z}}{2}
\bar{\chi}^{ll}(\pi+\alpha)+\delta_{l,z}\bar{\chi}^{ll}(\pi)\right]
\right\}.
\end{equation}
We expect $\bar{\chi}^{ik}(0)$ to become small at low $T$.\cite{RENARD87} 
This follows from 
the fact that it must vanish exponentially in the limit where 
rotational symmetry around the $z$-axis is exact. In this case we 
expect that the groundstate has $S^z_T=0$ and that there is a 
finite gap, $\Delta_x=\Delta_y$ to the lowest state of non-zero $S^z_T$. 
Thus, at low $T$, $\bar{\chi}^{zz}(0)\propto e^{-\Delta_x/T}$. The fact 
that in NENP, $\Delta_x\approx \Delta_y$ suggests that 
this symmetry is broken only by a small amount. This small 
symmetry breaking is presumably due to the $E^x$ term in Eq. (\ref{Ham0}) 
and the DM interaction. As pointed out in Ref. \onlinecite{RENARD87}, the 
$E^x$-term leads to a splitting of the gaps of first order in $E^x$ but 
a $T=0$ uniform susceptibility of second order in $E^x$. This suggests 
that $\bar{\chi}^{zz}(0)/\bar{\chi}^{xx}(0)$ should be of order 
$(2/15)^2=.018$. 
This estimate was confirmed by an explicit calculation using the 
Ginzburg-Landau field theory, reviewed above. On the other hand, 
the experiment obtained a value for this ratio of about .3. While 
it is possible that this just reflects errors in 
this rough estimate and in the detailed mean field  calculation\cite{RENARD87} 
which confirmed it, it seems more likely that another explanation is required. 
The explanation could reside in impurity effects or difficulties in 
separating the spin susceptibility from the diamagnetic contribution. 
However, later experiments\cite{AVENEL92} at lower $T$ suggest that the 
impurity 
contribution doesn't set in until considerably lower $T$ and 
that the data over the temperature range $T>1.7K$ may be dominated 
by the signal from the pure system. Thus we are led to consider 
the possibility that this discrepancy may be intrinsic. In this 
case, the obvious candidate is to include the staggered and uniform 
DM contributions and the staggered $g$-tensor at low T.
But at high $T$, both $\bar{\chi}^{ik}(0)$ and $\bar{\chi}^{ik}(\pi)$
go to  $2\delta^{ik}/(3T)$. So the staggered contribution to the 
susceptibility is suppressed by the small factors of $A^2$ and can be 
dropped. ( Since the staggered $g$-tensor is proportional to the difference
of $g$-tensor ($g_{\xi}-g_{\zeta}$), one may think dropping these terms 
will eventually affect the staggered $g$-tensor much. We actually did 
the calculation by keeping these terms and found the result changes very 
little.)
We also expect the relatively small crystal field Hamiltonian itself 
to become unimportant at high $T$. 
In this limit we have:
\begin{equation}
\chi^{ii}
={\partial^2 F\over \partial h_i^2}
= \mu_{B}^{2}\sum_kg^u_{ik}g^u_{ik} \bar{\chi},\end{equation}
where $\bar{\chi} \to 2/(3T)$ at large $T$.
Using the chain rule, the experimental measurements of susceptibility 
data in the crystallographic co-ordinate system at high $T$ thus give 
us, approximately, the following results for the $g$-tensor in the 
(x, y, z) co-ordinate system:
\begin{eqnarray}
g_{x}^2\cos^2{\phi}+g_{y}^2\sin^2{\phi}&=&(2.23)^2 \nonumber \\
g_{x}^2\sin^2{\phi}+g_{y}^2\cos^2{\phi}&=&(2.21)^2 \nonumber \\
g_{z}^2&=&(2.15)^2.
\end{eqnarray}
Setting $\theta=10^{\circ}$, $\phi=58^{\circ}$, we have from Eq. (\ref{gu}) 
\begin{equation}
g_{\xi}=2.20, \ \ g_{\eta}=2.24, \ \ g_{\zeta}= 2.15.
\end{equation}
Thus the uniform and staggered $g$-tensor in the $(x,y,z)$ coordinate 
system are, from Eqs. (\ref{gu}) and (\ref{gs}):
\begin{equation} 
\mathbf{g}^{u}
=\left( \begin{array}{ccc}
2.20 & 0 & 0 \\
0 & 2.24 & 0 \\
0 & 0 & 2.15
\end{array} \right )
,
\end{equation}
\begin{equation} 
\mathbf{g}^{s}
=\left( \begin{array}{ccc}
0 & 0 & 0.008 \\
0 & 0 & 0 \\
0.008 & 0 & 0
\end{array} \right )
,
\end{equation}
and matrix $\mathbf{A}$ can be derived from Eq. (\ref{hstag}).

For the GL model, from Eq. (\ref{GLM}) we can see there appear two extra 
terms when a magnetic field is applied to the system:
\begin{equation}
\delta H=-\int dz \mu_{B} \left[
\sum_{iklmn} h_{i} {\mathbf g}^{u}_{ik} {\cal R}_{kl}(\alpha \cdot z)
\epsilon_{lmn}\phi_{m} \Pi_{n}
+\sum_{ik}h^{s}_{i} {\cal R}_{ik}(\alpha \cdot z) \rho_{k}\phi_{k}\right].
\end{equation}
Now we use the free field approximation ($\lambda$=0) to calculate the  
susceptibilities of the spin chain.
Expanding $\phi_{i}$ and $\Pi_{i}$ (i=x, y, z) 
in terms of annihilation and creation operators:
\begin{equation}
\phi_{i}=\sum_{k}\sqrt{\frac{v_i}{2L\omega_{ik}}} 
\{exp[-i(\omega_{ik}t-kx)]a_{ik}+h.c.\}, 
\end{equation}
\begin{equation}
\Pi_{i}=\sum_{k}\sqrt{\frac{1}{2v_iL\omega_{ik}}}(-i\omega_{ik}) 
\{exp[-i(\omega_{ik}t-kx)]a_{ik}-h.c.\}, 
\end{equation}
where L is the number of spins and 
\begin{equation}
\omega_{ik}^{2}=\Delta_{i}^{2}+v^{2}k^{2}. 
\end{equation}
If there is no external field, the Hamiltonian becomes
\begin{equation}
H_{0}=\sum_{ik}\omega_{ik}(a^{+}_{ik} a_{ik} + \frac{1}{2}). 
\end{equation}
Looking at $\delta$H as a small term, we use first order
perturbation theory for eigenstate $\mid$n$>$, i.e,
$$|n> \to |n>-\sum_{m}\frac{|m><m| \delta H |n>}{E_{m}-E_{n}},$$ 
we have the finite-T formula for susceptibility 
\begin{equation}
\chi=\frac{2}{Z} \sum_{n,m}\frac{e^{-\beta E_{n}}\mid <n|(\delta H/h)|m>
\mid^{2}}{E_{m}-E_{n}}. 
\label{chi_cal}
\end{equation}
Define reduced uniform and staggered susceptibilities $\bar{\chi_u}$, 
$\bar{\chi_s}$ at momentum q ($q<<\pi$) as:
\begin{eqnarray}
\bar{\chi}_u^{ik}(q)&=&\frac{1}{L}\sum_{j_1 j_2}\int_0^\beta d\tau 
e^{q(j_{1}-j_{2})}<S^i_{j_1}(\tau )S^k_{j_2}(0)>
\nonumber \\
\bar{\chi}_s^{ik}(q)&=&\frac{1}{L}\sum_{j_1 j_2}\int_0^\beta d\tau 
e^{(\pi+q)(j_{1}-j_{2})}
<S^i_{j_1}(\tau)S^k_{j_2}(0)>.
\label{sus_dim}
\end{eqnarray}
From Eq. (\ref{chi_cal}) we can calculate them in the GL model
as follows:
\begin{equation}
\bar{\chi}_u^{ii}(q)
=\frac{1}{2}\int (\frac{dk}{2\pi}) 
\frac{1}{\omega_{l}\omega'_{m}}
\left[ (1+n_{l}+n'_{m})\frac{(\omega_{l}-\omega'_{m})^{2}}
{\omega_{l}+\omega'_{m}}
+(n_{l}-n'_{m})\frac{(\omega_{l}+\omega'_{m})^{2}}{\omega'_{m}-\omega_{l}}
\right] \ \ (i\neq l \neq m)
\label{sus_dim_ii}
\end{equation}
and 
\begin{equation}
\bar{\chi}_s^{ii}(q)=\frac{\rho_{i}^2v}{\Delta_{i}^2+v^2q^2}.
\label{sus_s_dim_ii}
\end{equation}
In Eq. (\ref{sus_dim_ii}),
$n_{ik}$ is the bosonic occupation number 
\begin{equation}
n_{ik}=\frac{1}{[exp(\omega_{ik}/T)-1]}.
\end{equation}
and 
\begin{equation}
k'=-k-q, \ \ \omega_i=\omega_{ik}, \ \ n_{i}=n_{ik}, 
\ \ \omega'_i=\omega_{ik'}, \ \ n'_{i}=n_{ik'}.   
\end{equation}
When the magnetic field is applied along b-axis, the uniform DM interaction 
will shift the staggered susceptibility by momentum $\alpha$, but will have
no effect on uniform susceptibility. So from Eq. (\ref{sus}), the 
susceptibility along b-axis ($\vec{h}=h\hat{b}$) can be written as 
\begin{equation}
\chi^{b}=(g_{z}\mu_{B})^{2}\bar{\chi}_{u}^{zz}(0)
+\frac{1}{2}\left[(\mathbf{A}_{13}\mu_{B})^2+(\mathbf{A}_{23}\mu_{B})^2\right]
\left[\bar{\chi}_{s}^{xx}(\alpha)+\bar{\chi}_{s}^{yy}(\alpha)\right].
\label{chi_b}
\end{equation}
Similarly, we can calculate the susceptibility along the a-axis
($\vec{h}= h\hat{a}=h[(\cos{58^{\circ}})\hat{x}
-(\sin{58^{\circ}})\hat{y}]$).
Now the only effect of uniform DM interaction is shifting the uniform 
susceptibility by momentum $\alpha $, so we have
\begin{equation}
\chi^{a}=\frac{1}{2}[(g_{x}\mu_{B}\cos{58^{\circ}})^{2}+
         g_{y}\mu_{B}\sin{58^{\circ}})^{2}]
         [\bar{\chi}_u^{xx}(\alpha)+\bar{\chi}_u^{yy}(\alpha)]
         +\mu_{B}^{2}(\mathbf{A}_{31}\cos{58^{\circ}}
         -\mathbf{A}_{32}\sin{58^{\circ}})^{2} \bar{\chi}_s^{zz}(0)
\end{equation}
Up to now, we have ignored the velocity differences. 
The $k$-integral converges at $k\to \infty$ so that it 
is not necessary to introduce an ultra-violet cut-off.  Of course 
a physical cut-off (the lattice spacing) exists in the spin 
chain but, in the approximation $\Delta <<J$, including this 
effect makes only small corrections. On the other hand, 
taking into account the velocity differences (according to 
Ref.\onlinecite{SORENSEN94}, $v_x=v_y=121K$, $v_z=114K$), the integrals 
diverge logarithmically at large $k$. This implies stronger 
dependence on the details of the dispersion relation at 
larger $k$ and the ultra-violet cut-off. However, for the 
small velocity difference in NENP, we find the susceptibility has
very weak cut-off dependence. Changing the cut-off from 
$\pi $ to $100\pi$ only changes $\chi^a$ by about 1\%. 
We just simply ignore this velocity difference.

\begin{figure}[h]
\centerline{
\epsfysize=0.50\textwidth{{\epsfbox{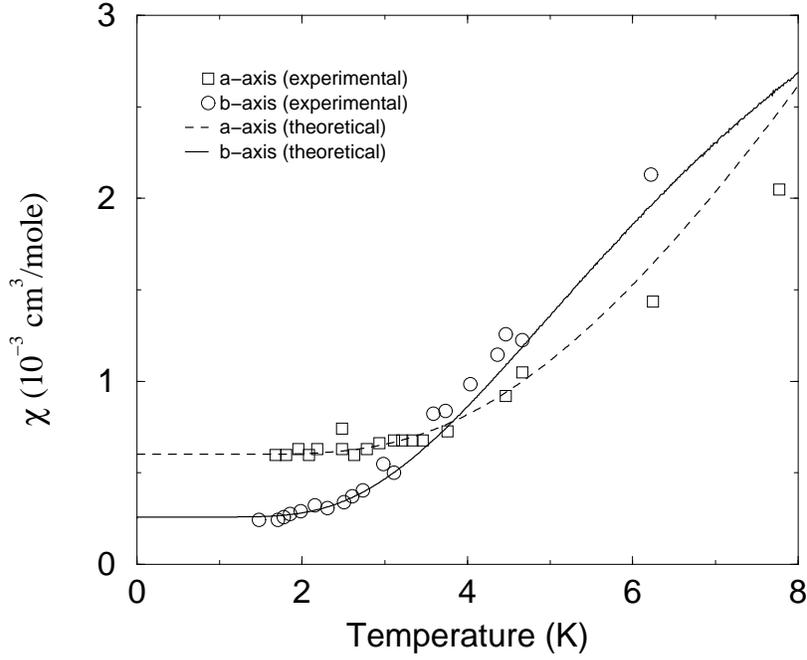}}}
}
\caption{Measured susceptibility vs the GL model prediction.}
\label{boson_tot}
\end{figure}

The staggered and uniform DM vectors are free parameters which can be 
chosen arbitrarily. We take
\begin{equation}
\frac{D_{x}}{2J}=-0.008, \ \ \frac{D_{y}}{2J}=-0.02, \ \ 
\frac{D_{z}}{2J}=0.04
\end{equation}
to fit the experimental data. We plot GL model results and experimental 
data in Fig. \ref{boson_tot}. The agreement is quite good at low $T$ where 
the field theory approximations are expected to work. 

\subsection{Susceptibility of Fermion Model: uniform and staggered}
\begin{figure}[h]
\centerline{
\epsfysize=0.50\textwidth{{\epsfbox{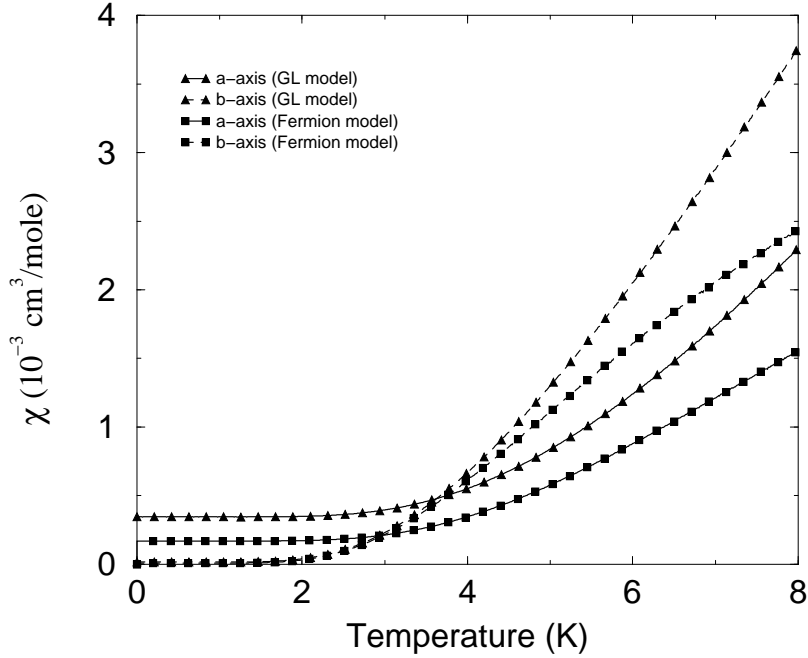}}}
}
\caption{The comparison of the uniform 
susceptibility of GL model ($D_{z}=0$) and that of fermion model ($D_{z}=0$).}
\label{com_u}
\end{figure}

For the fermion model, from Eq. (\ref{Tsvelik}) an extra term appears
when a uniform effective field is applied to the system: 
\begin{equation}
\delta H=-\int dz \left[\mu_{B}\sum_{klmnp}ih_{k}{\mathbf g}^{u}_{kl}
         {\cal R}_{lm}(\alpha\cdot z)\epsilon_{mnp}\bar{\chi}_{n}
         \gamma_{0}\chi_{p}\right].
\end{equation}
We can also use Eq. (\ref{chi_cal}) to compute the finite-T uniform 
susceptibility of this model. Let us set $D_{z}=0$ first, i.e., no uniform
DM interaction. Then $\cal R$ will be identity matrix.
We can easily get
\begin{equation}
\chi_{F}^{b}=(g_{z}\mu_{B})^{2}\bar{\chi}_{u,F}^{zz}(0)
\end{equation}
where in fermion model
{\small
\begin{equation}
\bar{\chi}_{u,F}^{ii}(q)=\int (\frac{dk}{2\pi}) 
\frac{1}{\omega_{l}\omega'_{m}}
\left[(1-n_{l,F}-n'_{m,F})\frac{\omega_{l}\omega'_{m}-|k| \cdot |k'|
-\Delta_l\Delta_m}{\omega_{l}+\omega'_{m}}
+(n_{l,F}-n'_{m,F})\frac{\omega_{l}\omega'_{m}+|k| \cdot |k'|
+\Delta_l\Delta_m}{\omega'_{m}-\omega_{l}}\right]
\ \ (i\neq l \neq m) \ .         
\label{chi_b_f}
\end{equation}}
Here $n_{i,F}$ is the fermionic occupation number 
\begin{equation}
n_{ik,F}=\frac{1}{[exp(\omega_{ik}/T)+1]}.
\end{equation}
and 
\begin{equation}
k'=-k-q, \ \ \omega'_i=\omega_{ik'}, \ \ n_{i,F}=n_{ik,F}, 
\ \ n'_{i,F}=n_{ik',F}.   
\end{equation}
Similarly,
\begin{equation}
\chi_{F}^{a}=(g_{x}\mu_{B}\cos{58^{\circ}})^{2}\bar{\chi}_{u,F}^{xx}(0)
        +(g_{y}\mu_{B}\sin{58^{\circ}})^{2}\bar{\chi}_{u,F}^{yy}(0)
        \label{chi_a_f}.
\end{equation}
The comparison of the uniform 
susceptibility of bosonic GL model ($D_{z}=0$) and that of fermion 
model ($D_{z}=0$) is shown in Fig. \ref{com_u}.
The fermion model results
are qualitatively similar to the bosonic results for the uniform
susceptibility although the $T=0$ value for $\chi^a$ is smaller by about a
factor of 50\% in the fermionic model. This makes the agreement with the
experimental data considerably worse before inclusion of staggered $g$-tensor
and DM interaction.   

\begin{figure}[h]
\centerline{
\epsfysize=0.50\textwidth{{\epsfbox{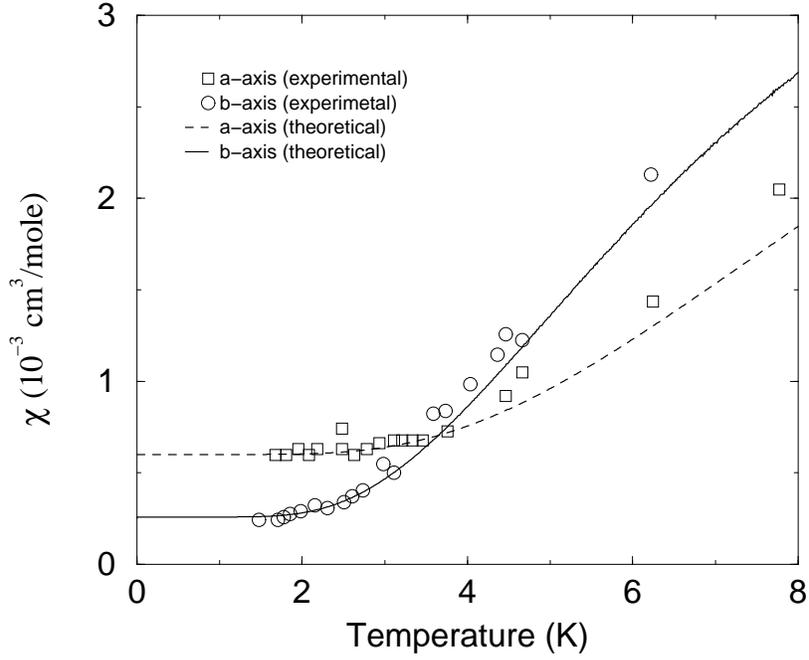}}}
}
\caption{Measured susceptibility vs the fermion model prediction.}
\label{fermion_tot}
\end{figure}

Similar to the GL model, we can also
include uniform DM interaction into susceptibility calculation. 
But as we discussed in Sec. IIIA, we don't know how to treat the staggered
effective field. If we include uniform DM interaction and simply
take the staggered contribution of GL model as that of fermionic model,
the total susceptibilities can be calculated and are plotted in
Fig. \ref{fermion_tot}.
The DM vectors are chosen as
\begin{equation}
\frac{D_{x}}{2J}=-0.005, \ \ \frac{D_{y}}{2J}=-0.03, \ \ 
\frac{D_{z}}{2J}=0.07
\end{equation}
to fit the experimental data. 
The agreement is roughly as good as GL model. 

\section{Discussion of Other Experimental Data and 
Other Theoretical Approaches}
Other experimental anomalies requiring staggered $g$-tensor and DM 
interaction for their explanation occur in the Knight shift 
and field-dependent gaps. Ignoring these small perturbations, 
the fermionic model is fairly successful at explaining 
the field-dependent gaps.\cite{TSVELIK90}  The GL model is accurate 
at low fields but captures 
only the qualitative features at higher fields where the 
Zeeman energy is of order the gap.\cite{AFFLECK90} We find that including 
the staggered $g$-tensor and staggered DM interaction 
does not significantly improve 
the agreement in the case of the GL model. Including the 
effect of the uniform DM interaction on the 
field dependent gaps is a fairly difficult problem
even in the GL model approximation since 
it requires a non-linear treatment of 
a slowly rotating field, and we do not attempt it here.
As remarked above there appears to be no simple way of 
including the staggered field in the fermion model, thus 
precluding a calculation of its effects on field-dependent 
gaps in that model. The field-dependent Knight
shift,\cite{CHIBA91} presents similar 
calculational difficulties using GL or fermion model.

There is another low-energy effective field theory model that 
has been applied to NENP. 
This model was proposed by Mitra and Halperin.\cite{MITRA94}
The Hamiltonian is the following
\begin{equation}
H=\int dz \left\{ \sum_{i} \left[\frac{v}{2}{\Pi_{i}}^{2}
+\frac{v}{2}\left(\frac{\partial \phi_i}{\partial z}\right)^{2}+
\frac{\Delta_{i}^{2}}{2v}\phi_{i}^{2}\right]
-\mu_{B}\sum_{iklm}h_{i}{\mathbf g}^{u}_{ik}\epsilon_{klm}
\sqrt{\frac{\Delta_{l}}{\Delta_{m}}}\phi_{l}\Pi_{m}
-\mu_{B}\sum_{i}h^{s}_{i}\rho_{i}\phi_{i}
+\lambda \left.\vec{\phi}\,\right.^4
\right\}.
\end{equation}
In this 
case we also ignore the small 
velocity difference which can be shown to have a negligible 
effect, as in the GL model. This model differs 
from the standard GL theory of Eq. (\ref{GLM}) by the 
factors of $\sqrt{\frac{\Delta_l}{\Delta_m}}$ in the coupling to 
the magnetic field. 
These factors were introduced in Ref. (\onlinecite{MITRA94}),
in order to obtain field dependent gaps which are 
the essentially the same as in the fermionic model and hence 
agree much better with experiment. 

In mean field approximation, the uniform susceptibility is:
\begin{equation}
\chi^{b}=\frac{1}{2}(g_{z}\mu_{B})^{2}\int (\frac{dk}{2\pi}) 
\frac{1}{\omega_{x}\omega_{y}}
\left[(1+n_{x}+n_{y})\frac{\left(\omega_{x}\sqrt{\frac{\Delta_y}{\Delta_x}}-
\omega_{y}\sqrt{\frac{\Delta_x}{\Delta_y}}\right)^{2}}{\omega_{x}+\omega_{y}}
+(n_{x}-n_{y})\frac{\left(\omega_{x}\sqrt{\frac{\Delta_y}{\Delta_x}}
+\omega_{y}\sqrt{\frac{\Delta_x}{\Delta_y}}\right)^{2}}{\omega_{y}-\omega_{x}}
\right]
\end{equation}   
and similarly for $\chi^a$.  Note that the $k$-integral 
does not converges in the ultraviolet, even ignoring velocity 
differences,  since the integrand 
behaves as $(\sqrt{\frac{\Delta_y}{\Delta_x}}-
\sqrt{\frac{\Delta_x}{\Delta_y}})^2/v|k|$
at large $k$. Thus the result will be more sensitive to the details 
of the ultra-violet cut-off than for the other models considered above.
Choosing a cut-off, $|k|<\pi$, and ignoring the staggered field 
gives  a value for 
$\chi^a(T=0)$ which is about twice as large as that observed experimentally. 
Including the staggered field raises the theoretical result still 
higher making the agreement worse.  Furthermore, choosing 
the arbitrary cut-off to be $10\pi$, increases $\chi^a(T=0)$ by 
a factor of about 2, also worsening the agreement.  

Sieling et al.\cite{SIELING00}, using the Lancz\"{o}s algorithm and 
the density matrix renormalization group technique, studied 
the field-induced gaps, for a field in the z-direction, using 
a model containing an alternating field and alternating 
as well as uniform crystal field terms. Independent rotation 
matrices were assumed for these two types of alternating 
terms, rather than assuming that both $g$-tensor and 
crystal field tensors are diagonal in the $(\xi, \zeta ,\eta )$ 
co-ordinate system, as seems likely. The DM interaction was not 
included. More numerical work of this type, including 
the DM interaction (and perhaps also the staggered crystal 
field terms) and considering the other field direction 
and the susceptibilities is needed to 
determine accurate values of the staggered $g$-tensor and DM 
interactions.  

We would like to thank M. El-Batanouny for 
asking a question which prompted this investigation 
and for very helpful discussions on 
crystal symmetry. This research was supported by NSF grant No. 
DMR 02-03159.

\end{document}